%% file: demo_paper.tex
\documentclass[conference]{IEEEtran}
\IEEEoverridecommandlockouts
\usepackage{cite}
\usepackage{amsmath,amssymb,amsfonts}
\usepackage{algorithmic}
\usepackage{hyperref}
\usepackage{graphicx}
\usepackage{textcomp}
\usepackage{xcolor}
\def\BibTeX{{\rm B\kern-.05em{\sc i\kern-.025em b}\kern-.08em
    T\kern-.1667em\lower.7ex\hbox{E}\kern-.125emX}}
\begin{document}

\title{Demo - Zelig: Customizable Blockchain Simulator
\thanks{Accepted for presentation at the International Symposium on Reliable Distributed Systems (SRDS 2021).}
}

% \author{Anonymous Authors}

\author{\IEEEauthorblockN{Ege Erdoğan}
\IEEEauthorblockA{
\textit{Koç University}\\
Istanbul, Turkey \\
eerdogan17@ku.edu.tr}
\and
\IEEEauthorblockN{Can Arda Aydın}
\IEEEauthorblockA{
\textit{Koç University}\\
Istanbul, Turkey \\
canaydin17@ku.edu.tr}
\and
\IEEEauthorblockN{Öznur Özkasap}
\IEEEauthorblockA{
\textit{Koç University}\\
Istanbul, Turkey \\
oozkasap@ku.edu.tr}
\and
\IEEEauthorblockN{Waris Gill}
\IEEEauthorblockA{
\textit{Koç University}\\
Istanbul, Turkey \\
wgill18@ku.edu.tr}
}

\maketitle

\begin{abstract}
As blockchain-based systems see wider adoption, it becomes increasingly critical to ensure their reliability, security, and efficiency. Running simulations is an effective method of gaining insights on the existing systems and analyzing potential improvements. However, many of the existing blockchain simulators have various shortcomings that yield them insufficient for a wide range of scenarios. In this demo paper, we present Zelig: our blockchain simulator designed with the main goals of customizability and extensibility. To the best of our knowledge, Zelig is the only blockchain simulator that enables simulating custom network topologies without modifying the simulator code. We explain our simulator design, validate  via experimental analysis against the real-world Bitcoin network, and highlight potential use cases.  
\end{abstract}


\section{Introduction}

Introduced in 2008 as the building block of Bitcoin \cite{Nakamoto_bitcoin}, a blockchain essentially enables a dynamic set of distrusting parties to reach consensus on a history of transactions without relying on a trusted third party. Although the hype around blockchains seems to overshadow their actual uses, novel blockchain-based systems have been proposed in different problem domains such as peer-to-peer energy trading \cite{Computer2020,TII2021}, electronic voting \cite{YANG2020859}, supply chain management \cite{sc_dutta2020blockchain}, and finance \cite{Nakamoto_bitcoin, wood2014ethereum}.

% As the saying goes, with great power comes great responsibility. 
If blockchain-based systems are to be used for purposes that are highly crucial for the well-being of a society, it is of critical importance to ensure that those systems are secure, efficient, and reliable.
%
%For this purpose, methods and tools that enable evaluation of blockchain systems can prove to be valuable. 
Monitoring, benchmarking, experimental analysis, and running simulations are the main methods that allow empirical analysis of blockchain systems \cite{fan_performance_2020}. Among these methods, running simulations stands out as an accessible and low-cost alternative. While the other methods require direct access to the system under evaluation, simulations can run independently and cover a wide range of scenarios in a short amount of time. 

However, most existing simulators have various shortcomings that make them unfit for evaluating a wide range of scenarios \cite{paulavicius_systematic_2021}. Since most of them are designed with a specific but limited purpose (e.g. to demonstrate an attack \cite{eyal_majority_2013}, or test certain properties of the system \cite{gervais_security_2016}), they make unrealistic assumptions to simplify the layers that are irrelevant for their purposes. Furthermore, most simulators present little customization opportunities beyond parameter-tuning \cite{paulavicius_systematic_2021}. 
A well-designed and easily customizable simulator would not only be an achievement in itself but also act as a catalyst for further research. Such a simulator could be used to test blockchain-based systems' reliability, efficiency and security in varying network conditions, or in the presence of adversaries. 

\textbf{Contributions.} The key contribution in this paper is Zelig\footnote{The name \textit{Zelig} refers to the 1983 Woody Allen film, telling the story of Leonard Zelig, who assumes the physical properties of anyone he comes into contact with to fit in.}, our blockchain simulation framework, designed with the main goals of customizability and extensibility, implemented in Python. By supplying the users with ready-made primitives and ample customization and extension opportunities on various levels through a well-defined interface, we aim to streamline further research that would otherwise require building a simulator from scratch. Our implementation currently consists of a \textit{core simulator} that captures the main elements of a blockchain system, and a Bitcoin simulator that can be used as a blueprint for similar simulations. To the best of our knowledge, Zelig is the only blockchain simulator that allows users to simulate custom network topologies without modifying the simulator code. 

Our demonstrations would mainly involve running Bitcoin simulations under different scenarios, one of which is the selfish mining attack \cite{eyal_majority_2013}. We would then analyze its effectiveness with different settings to demonstrate how Zelig simulator would benefit such research. In the rest of the paper, we describe Zelig's design (Section \ref{design}), validate it experimentally against the real-world Bitcoin network (Section \ref{validation}), and describe sample use cases (Section \ref{use}). The source code for Zelig can be found at \href{https://github.com/ZeligSim/Zelig}{https://github.com/ZeligSim/Zelig}.

\section{Zelig Simulator Design} \label{design}

\subsection{Core Simulator}

\begin{figure}[t!]
    \centering
    \includegraphics[width=0.49\textwidth]{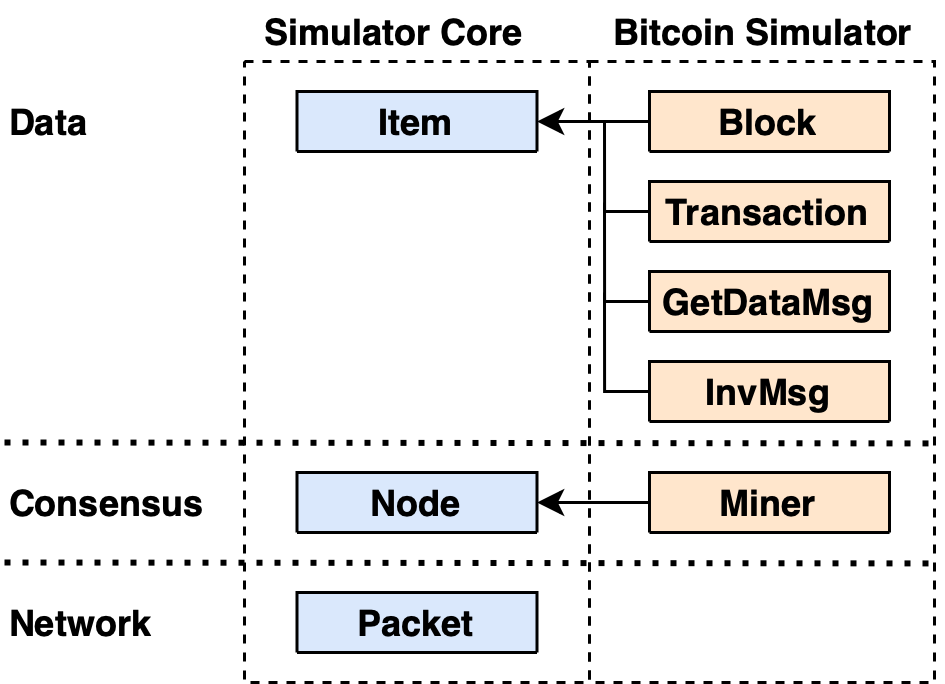}
    \caption{Class diagram of Zelig's core simulator and the extended Bitcoin simulator with the system layer they correspond to (data, consensus, and network layers). Each class in the simulator core corresponds to a specific layer, and the Bitcoin simulator extends them.}
    \label{fig:uml}
\end{figure}

Attempting to capture the main elements of a blockchain system in a modular way, the core simulator consists of three classes: \textit{Item}, \textit{Node}, and \textit{Packet}. As displayed in Figure \ref{fig:uml}, each class corresponds to a specific layer of a blockchain system. We follow the layering approach that divides a blockchain system into the layers of application, execution, data, consensus, and network  \cite{fan_performance_2020}. Zelig implementation consists of:
%Since the application and execution layers are beyond our simulator's scope, we focus on the three remaining layers:
\begin{itemize}
    \item The \textit{data layer} corresponds to the data (blocks, transactions) stored on the blockchain and the method (hashing, Merkle  trees) used for storage.
    \item The \textit{consensus layer} involves the protocol for the nodes reach agreement on the system state (e.g. proof-of-work, proof-of-stake, PBFT). 
    \item The \textit{network layer} corresponds to the P2P network that the nodes communicate over. 
\end{itemize}

Zelig's three core classes are further detailed as follows:
\begin{itemize}
    \item The \textit{Item} class corresponds to the data that is being exchanged between the nodes and stored on the blockchain. It can be extended to represent objects such as blocks or protocol-specific messages.
    \item The \textit{Node} class represents the participants in the blockchain network, and is the main entry point for running simulations. At each simulation step, all nodes' \texttt{step()} methods are called. In each call of its \texttt{step()} method, a node increments its timestamp by one, obtains the items corresponding to that timestamp from its inbox, and performs the necessary actions on those items. The order in which the nodes' \texttt{step()} methods are called does not affect the simulation result, and thus the simulations can be easily parallelized.
    \item The \textit{Packet} class is a wrapper for \textit{Item}s that abstracts away the network layer logic from the data layer logic. Each \textit{Packet} has a payload (an \textit{Item}), and a timestamp value at which the packet would be processed by its recipient.  
\end{itemize}

\subsection{Simulating Bitcoin}
% \footnote{Although we describe the implementation as simulating \textit{Bitcoin}, it can easily be configured to simulate similar altcoins such as Litecoin and Dogecoin that only require changes to the block interval values.}}

To demonstrate how the core simulator can be extended, we implemented a Bitcoin simulator on top of the core simulator, which can in turn be extended to add different behaviours. Details of the three layers of the Bitcoin simulator are provided next.

\subsubsection{Data Layer}

We simulate the Bitcoin data layer by extending the \textit{Item} class to represent \textit{Inv} and \textit{GetData} messages, as well as \textit{Blocks} and \textit{Transactions}. Transactions are implemented in a simplified way as of this writing, and we plan to supply a more realistic implementation by analyzing the real-world transaction data in detail.
% We should note that the transaction implementation is in a preliminary stage as of this writing.

In addition to the fields of its parent \textit{Item}, a \textit{Block} object maintains a pointer to its preceding block in the blockchain. When a block is created, its transaction count is sampled from a normal distribution with mean $2104.72$ and standard deviation $236.63$, and the size of each transaction from a normal distribution with mean $615.32$ bytes and standard deviation $89.43$. We extracted these values by analyzing the last 365 days' transaction data obtained from \cite{tx_site}.

\subsubsection{Consensus Layer}

The consensus layer is exclusively implemented in the \textit{Miner} class, addressing block generation and fork resolution problems. Each miner has a fixed mining power that determines its probability of generating a block at each step. 

As it would be infeasible to implement the actual proof-of-work mechanism, nodes generate a block with probability
\begin{equation}
    P = \frac{\text{NodeMinePower}}{\text{BlockInterval} \cdot \text{TotalMinePower}}.
\end{equation}
This ensures that miners generate blocks in proportion with their mining powers, and that the expected time between two blocks is the specified block interval. Since nodes generate blocks independently, forks can occur. Miners resolve forks by mining on top of the longest chain determined by its height, since all blocks have the same difficulty for now. 

\subsubsection{Network Layer}

The network layer consists of the \textit{Packet} class described above. A packet's transmission delay depends on its payload's size and the geographic regions of its sender and recipient. Each region has fixed upload and download values (MB/s), and each pair of regions has a fixed latency value (ms). Then, considering regions $A$ and $B$ with latency $L_{AB}$, download bandwidths $D_A, D_B$, and upload bandwidths $U_A, U_B$, the network delay for a message with size $S$ from region $A$ to region $B$ is computed as 
\begin{equation}
    \text{Delay} = L_{AB} + \frac{S}{\text{min}\{U_A, D_B\}}.
\end{equation}

\input{tables/exp_setups}

\input{tables/results}

\section{Experimental Validation} \label{validation}

In this section, we compare the preliminary experimental results obtained from Zelig with the real-world Bitcoin network. In these experiments, we consider nine world wide regions
% \footnote{With the details omitted for the sake of brevity the regions are China, America, Russia, Kazakhstan, Malaysia, Canada, Germany, Venezuela, and Norway.}
that make up more than $90\%$ of the total mining power. We obtain the regions' bandwidth, node count, and mining power values from \cite{bandwidth_site, nodes_site}, and \cite{mine_site} respectively. The performance metrics of interest that we use for comparison are:

\begin{itemize}
    \item \textbf{Average block interval} ($\Delta_B$): Mean time between two blocks on the main blockchain.
    \item \textbf{Stale block rate} ($r_s$): Percent share of \textit{stale} blocks (mined but not included in the main chain) among all blocks.
    \item \textbf{Block propagation delays} ($d_p$): The time it takes for blocks to reach a certain share ($p\%$) of the nodes.
    \item \textbf{Transaction throughput} ($tps$): Number of transactions included in the main chain per second.
\end{itemize}

Table \ref{tab:exp_setups} describes the experimental setups, and Table \ref{tab:results} compares our results with the real-world Bitcoin network. We obtain stale block rate ($r_s$) and block propagation delay values ($d_p$) for the real-world network from \cite{paulavicius_systematic_2021}, average block intervals ($\Delta_B$) from \cite{block_interval_site}, and transaction throughput values ($tps$) from \cite{tx_site}.

Overall, the results are reasonably close to the real-world network. We believe we can obtain more accurate results  once we implement recent developments such as compact block relay \cite{corallo_2020} that reduce propagation delays and thus the stale block rate.

%\section{Sample Use Cases} \label{use}
\section{Use Cases and Future Work} \label{use}

In this section, we highlight two potential use cases in which Zelig can be customized in its current state to simulate additional interesting scenarios, and summarize future work.

\textbf{Simulating Different Attack Scenarios.} Attacks such as selfish mining \cite{eyal_majority_2013} can pose serious risks to a blockchain system's security. It is infeasible to test such attacks on the real-world network, and implementing a simulation environment from scratch can be a cumbersome task for different research works. By providing an easy-to-use test bed for such attacks, Zelig can be a valuable tool for conducting such research. In Zelig, the Bitcoin \textit{Miner} class can be extended without changing the rest of the simulator to implement different attack scenarios (e.g. selfish mining and eclipse attacks). 
%(e.g. selfish mining \cite{eyal_majority_2013}, eclipse attacks \cite{heilman_eclipse_nodate}). 
Testing the attack using Zelig and writing no more than the bare minimum attack code itself can speed up such research, and enable potential vulnerabilities to be discovered faster.

\textbf{Experimenting with Custom Network Topologies.} A relatively less explored area of public blockchains is the effect of the network topology on system behavior. In Zelig, different network topologies can be implemented by specifying a Python function that takes two nodes as input and returns \texttt{True} if the nodes are connected. For example, a ring topology with 10 nodes can be constructed by providing the following Python \texttt{lambda} expression when configuring the simulator: 
\begin{verbatim}
lambda n1, n2: abs(n1.id - n2.id) == 1 
                or abs(n1.id - n2.id) == 9
\end{verbatim}
Since designing blockchain systems with well-defined network structures can provide potential performance benefits, Zelig has also potential to speed up such research by freeing the researchers from the burden of implementing their custom simulators. 

%\section{Future Work}

\textbf{Future Work.} We plan to extend the Zelig simulator by implementing further customization and extension opportunities, for instance by modelling an incentive mechanism for miners. Additional future directions would be bringing the Bitcoin simulator up to date with recent developments, and providing blueprint implementations for different platforms such as Ethereum \cite{wood2014ethereum}.

\bibliographystyle{ieeetr}
\bibliography{references}

\end{document}

%% file: tables/exp_setups.tex
\begin{table}[h!]
\centering
    \caption{Input parameters for the six experiments we have performed (M: million, Full: real-world node counts).}
    \begin{tabular}{|c|c|c|c|c|c|c|}
         \hline
         \textbf{Parameter} & \textbf{\#1} & \textbf{\#2} & \textbf{\#3} & \textbf{\#4} & \textbf{\#5} & \textbf{\#6} \\\hline 
         Nodes per region   & 1             & 10           & 1            & Full         & Full         & Full         \\ \hline
         Simulation steps   & 30M           & 30M          & 30M          & 15M          & 15M          & 15M  \\ \hline
         Connections per node &  2          & 2            & 2            & 32           & 8            & 4     \\ \hline
         Seconds per step   & 0.1           & 0.1          & 1            & 0.1          & 0.1          & 0.1  \\ \hline
    \end{tabular}
    \label{tab:exp_setups}
\end{table}

%% file: tables/results.tex
\begin{table*}[t!]
    \centering
    \caption{Comparison between the real-world Bitcoin metrics and the results obtained from Zelig.}
    \begin{tabular}{|c|c|c||c|c|c|c|c|c|}
         \hline
                                & \multicolumn{2}{|c||}{\textbf{Bitcoin}} & \multicolumn{6}{|c|}{\textbf{Simulation}} \\ \cline{2-9}
        \textbf{Metric}            & 2016     &  2020   & \#1    & \#2    & \#3    & \#4    & \#5    & \#6 \\\hline
        $r_s$ (\%)                      & 0.41     & 0.06    & 0.34   & 0.69   & 0.44   & 0.58   & 0.55   & 0.63   \\ \hline
        $\Delta_B$ (s)             & 9.69     & 10.05   & 9.96   & 10.04  & 10.19  & 9.94   & 10.17  & 10.51   \\ \hline
        $tps$                      & 2.87     & 3.48    & 3.53   & 3.52   & 3.52   & 3.55   & 3.47   & 3.36  \\ \hline
        $d_{50}$ (s)               & 8.7      & 0.5     & 2.49   & 4.51   & 5.07   & 2.25   & 3.14   & 4.06   \\ \hline
        $d_{90}$ (s)               & 17.3     & 3.3     & 4.35   & 6.29   & 8.82   & 2.77   & 3.62   & 4.6   \\ \hline
    \end{tabular}
    \vspace{-0.5cm}
    \label{tab:results}
\end{table*}